\documentclass[twocolumn,nofootinbib,prl,floatfix,preprintnumbers,amsmath,amssymb,groupedaddress,superscriptaddress]{revtex4}
\usepackage{mciteplus}
\usepackage{dsfont}
\usepackage[pdftex]{graphicx} 
\usepackage{epstopdf} 
\usepackage{slashed}

\newcommand{\definmath}[2] {\def#1{\ifmmode#2\else$#2$\fi}}
\definmath{\invfb}{\mathrm{fb}^{-1}}

\begin{document}   
\title{
Transverse masses and kinematic constraints:\\from the boundary to the crease
}
\date{\today}

\author{Alan J. Barr}
\email{a.barr@physics.ox.ac.uk}
\affiliation{Denys Wilkinson Building, Keble Road, Oxford, OX1 3RH, United Kingdom}

\author{Ben Gripaios}
\email{gripaios@cern.ch}  
\affiliation{CERN PH-TH, Geneva 23, 1211 Switzerland}

\author{Christopher G. Lester}
\email{lester@hep.phy.cam.ac.uk}
\affiliation{Cavendish Laboratory, Dept of Physics, JJ Thomson Avenue,
Cambridge, CB3 0HE, United Kingdom}

\preprint{Cavendish-HEP-09/16}
\preprint{CERN PH-TH/2009-158}
\begin{abstract} 
We re-examine the kinematic variable $m_{T2}$ and its relatives
in the light of recent work by Cheng and Han.
Their proof that $m_{T2}$ admits an equivalent, but implicit,
definition as the `boundary of the region of parent and daughter
masses that is kinematically consistent with the event hypothesis' is
far-reaching in its consequences.  We generalize their result both to
simpler cases ($m_T$, the transverse mass) and to more complex cases
($m_{TGen}$).  We further note that it is possible to re-cast many
existing and unpleasant proofs (e.g.~those relating to the existence
or properties of ``kink'' and ``crease'' structures in $m_{T2}$) into almost trivial
forms by using the alternative definition.  Not only does this allow
us to gain better understanding of those existing results, but it also
allows us to write down new (and more or less explicit) definitions of
(a) the variable that naturally generalizes $m_{T2}$ to the case in
which the parent or daughter particles are not identical, and (b) the
inverses of $m_T$ and $m_{T2}$ -- which may be useful if daughter
masses are known and bounds on parent masses are required.  We note
the implications that these results may have for future matrix-element
likelihood techniques.
\end{abstract}   
\maketitle 
\section{Introduction}
If a dark matter candidate is produced at the LHC, one of our experimental priorities will be to measure its mass. 
This is a non-trivial exercise, since dark matter is, by its very nature, invisible in the detectors. Thus, kinematic information is lost in each event. What is more, the difficulties are compounded by the fact that dark matter is invariably pair-produced.
Despite receiving a lot of attention in the recent literature, we still only have one method for measuring masses\footnote{Once the dynamics are known or postulated, one could hope to measure masses directly from the matrix element. We comment on this later.} in the case of pair decays which are identical, with each containing one invisible particle in the final state,\footnote{For decays with more than one invisible particle, see \cite{Barr:2009mx}.} but are otherwise arbitrary \cite{Cho:2007qv,Gripaios:2007is,Barr:2007hy,Cho:2007dh}.\footnote{In special cases, involving cascade decays and extra kinematic constraints, other methods are available 
\cite{Hinchliffe:1996iu,Paige:1996nx,Bachacou:1999zb,Hinchliffe:1999zc,atlasphystdr,Allanach:2000kt,Lester:2001zx,Nojiri:2003tu,Kawagoe:2004rz,Gjelsten:2004ki,Gjelsten:2005aw,Lester:2005je,Miller:2005zp,Lester:2006cf,Tovey:2008ui,Kersting:2008qn,Bian:2006xx,Ross:2007rm,Cheng:2008mg,Burns:2008va,Barr:2008ba,Nojiri:2008vq,Bisset:2008hm,Barr:2008hv,Serna:2008zk,Huang:2008ae,Cho:2008tj,Konar:2008ei,Cheng:2009fw,Burns:2009zi,Matchev:2009iw,Choi:2009hn,Costanzo:2009mq,Kersting:2009ne,Kang:2009sk,Webber:2009vm}.
An alternative method for the general case making use of the distribution of initial state radiation has also recently been proposed \cite{Papaefstathiou:2009hp}.
} The method is based on the $m_{T2}$ variable, introduced in \cite{Lester:1999tx,Barr:2003rg} and defined by
\begin{gather}
m_{T2} \equiv \mathrm{min}\ \mathrm{max} (m_T, m_T^\prime) .
\end{gather}
Here $m_T$ and $m_T^\prime$ are the transverse mass variables for the individual decays, introduced originally for measuring the mass of the $W$-boson \cite{Arnison:1983rp,Banner:1983jy}, and defined explicitly below. The `max' tells us to take the larger of these two variables. In the minimization, one is instructed to consider all possible partitions of the measured missing transverse momentum in the event between the two invisible particles, and to minimize with respect to partitions.

This {\em ad hoc} definition is designed to cope with the fact that only the sum of the transverse momenta of the two invisible particles can be inferred from the missing transverse momentum observed in a collision, whilst inheriting one desirable property from the usual transverse mass: it is bounded above by the mass of the parent particle. 
Unfortunately, $m_{T2}$ is still not an observable when the masses of the invisible daughters are unknown and non-negligible, because the usual transverse mass is a function of the invisible daughter mass. In the case of neutrinos, whose mass can be neglected, this is not a problem. But it certainly is a problem for dark matter candidates.

To cope with this obstacle, a futher {\em ad hoc} step was taken: consider $m_{T2}$ as a function of the unknown invisible daughter mass $m_i$: $m_{T2} (m_i)$. Now $m_{T2} (m_i)$ is an observable, albeit an observable function. That is to say, each detector event returns a function. Unfortunately, in taking this step, the boundedness property of $m_{T2}$ is lost: it is not true that $m_{T2} (m_i)$ is bounded above by the mass of the parent, for abitrary values of $m_i$.

What then are the properties of $m_{T2} (m_i)$? On an event-by-event basis, $m_{T2} (m_i)$ is simply a smooth function of $m_i$. But if one plots the envelope of curves coming from many events, one discovers that the maximal curve features a kink \cite{Cho:2007qv,Gripaios:2007is,Barr:2007hy,Cho:2007dh}. That is to say, it is continuous, but not differentiable, exactly at the point $m_i= \tilde{m_i}$, where $\tilde{m_i}$ is the true mass of the invisible daughter.  Moreover, since we already know that the maximal value of $m_{T2} (\tilde{m_i})$ is the (true) parent mass, $\tilde{m_0}$, we see that the kink has co-ordinates $(\tilde{m_i},\tilde{m_0})$; by identifying the location of the kink in an experiment, one may measure the masses of both the parent and the invisible daughter.

On reflection, this result, though pleasing, is somewhat mysterious. One started from an {\em ad hoc} definition of a transverse mass for identical pair decays, guaranteeing only the desirable property that $m_{T2} (\tilde{m_i}) \leq \tilde{m_0}$.  Moreover, generalization to $m_i \neq \tilde{m_i}$, in which the one desirable property contained in the definition is lost, gives rise to a strange kink behaviour. What does all of this mean?

Recently, Cheng and Han, gave an elegant interpretation of the function $m_{T2} (m_i)$ \cite{Cheng:2008hk}.\footnote{A similar interpretation was given without proof in \cite{Serna:2008zk}.} They showed that, for a given event, it defines the boundary of the region in the $(m_i, m_0)$ plane for which the various kinematic constraints, namely conservation of four-momentum and the mass shell constraints, admit a solution. By `admit a solution', one means that there exist real values of the unknown momenta, and real, non-negative values of the unknown energies, solving the constraints. The existence of such a solution for a given value of $(m_i, m_0)$ means that one cannot rule out the possibility that the true mass values, $(\tilde{m}_i, \tilde{m}_0)$, are given by $(m_i, m_0)$, on the basis of the information obtained from that event. 

The proof is very simple, though we refrain from repeating it here (we shall, in any case, give a proof for a more general case of non-identical pair decays in what follows). The beauty of the result is that it shows that the original {\em ad hoc} definition of $m_{T2} (\tilde{m_i})$, and its {\em ad hoc} extension to $m_{T2} (m_i)$, fortuitously give rise to a natural function, namely the function that defines the boundary of the allowed region in mass space, on an event-by-event basis. If one considers multiple events, the allowed region is restricted to the intersection of the allowed regions coming from each event. If one considers arbitrarily many events, one ends up with an extremal allowed region, and a corresponding extremal boundary. In the case of identical pair decays this extremal boundary is precisely that containing the kink identified in \cite{Cho:2007qv,Gripaios:2007is,Barr:2007hy,Cho:2007dh}.

In what follows, we would like to show that this re-interpretation of
$m_{T2} (m_i)$ in terms of a boundary is both general and
powerful. Firstly, we remark that it also applies to the simpler case
of the usual transverse mass variable for single-particle
decays. Secondly, we show how it can be used to give an almost trivial
derivation of the form of the kink curve for identical pair
decays. Thirdly, we show how it can be generalized to pair decays in
which either the parents, or the invisible daughters, or both, are not
identical, and have different masses. This generalization may be
practically useful, for example, in the case of squark-gluino
production in the context of supersymmetry, or in theories in which
more than one particle is stable on the length scale of a detector. In
these cases, the number of masses that are, {\em a priori}, unknown is
increased. Consequently, the mass space is higher-dimensional, and so
is the boundary of the allowed region that follows from applying the
kinematic constraints to an event. Nevertheless, the form of the
extremal boundary is easily obtained. For the case of distinct parents
(such as a squark and a gluino) decaying to a common LSP, the extremal
boundary forms a surface in the three-dimensional space parametrized
by the three unknown masses. The surface is, as we shall see, creased,
with various kink structures visible in two-dimensional
projections. Fourthly, we explain how observables introduced
previously to cope with combinatorics and upstream or initial state radiation can also be understood as generating the kinematic boundary.

We stress that our arguments are purely theoretical, and take no account of what might realistically be achieved in kinematic measurements at the LHC. Nevertheless,
we feel that, in order to properly understand and use variables like $m_{T2}$, it is important to know both how the variables behave under ideal conditions, as well as how this is modified in real situations. Although we shall not address the latter aspect here, we hope to make a useful contribution to the former.

As an example of how reality deviates from the ideal, it is clear that a real LHC data sample cannot saturate the true extremal boundary. Indeed, the extremal boundary corresponds to events in which the parent particles are infinitely boosted with respect to the laboratory frame \cite{Gripaios:2007is} by radiation upstream or in the initial state. 
Nevertheless, even a subset of events, such as those contained in a finite LHC data sample, defines a corresponding kinematic boundary. The arguments we give show that reconstruction of that boundary is the best that one can hope to achieve in the absence of additional kinematic or dynamic information, whether inferred or assumed. As another example, it is not yet clear how the boundary hypersurface we describe might best be reconstructed from LHC data. What one would like to do is to generate one-dimensional distributions of observables, which can then be fitted by Monte Carlo simulations. Towards the end of the paper we make a partial effort to address this, by discussing how such observables, related to the boundary hypersurface, may be derived.
  
Our notation is as follows. For a single particle decay, we consider a parent particle of mass $m_0$ decaying into an invisible daughter particle of mass $m_i$ and a system of visible daughter particles of invariant mass $m_v$. We write the four-momenta of particle $0$ by $p_0^\mu = (E_0, \mathbf{p}_0, q_0)$,
where $\mathbf{p}_0$ is transverse to the beam direction and $q_0$ is parallel to it. We denote the transverse energy-momentum by $\alpha_0 = (e_0, \mathbf{p}_0)$,
where the transverse energy is defined by $e_0^2 = \mathbf{p}_0^2 +m_0^2$. For pair decays, we use unprimed quantities for one decay and primed quantities for the other. We shall often need to distinguish between hypothesized values of the unknown masses of the parent and invisible daughter, and the true values; we denote the {\em latter} with tildes. 

Finally, to avoid confusion, we remark that we shall always illustrate our arguments with the special case where the visible daughter system contains a single, massless particle. Although theoretically the simplest case, this is probably the least favourable example from an experimentalist's viewpoint, since a kink is generated in this case only by events in which there is significant upstream transverse momentum. Nevertheless, our general arguments apply to arbitrary visible systems, including those which appear to be experimentally more favourable.
\section{Single particle decays and the Transverse Mass}
Let us first prove that for an event consisting of a single particle decay, the locus of the curve $m_0 = m_T (m_i)$ is equivalent to the boundary of the region in $(m_i,m_0)$ for which the kinematic constraints\footnote{Cheng and Han \cite{Cheng:2008hk} call these the minimal kinematic constraints, because one may have supplementary constraints in theories with cascade decays.} 
\begin{align}\label{kcsingle1}
p_i^2 &= m_i^2, \\ \label{kcsingle2}
p_0^2= (p_i + p_v)^2 &= m_0^2, \\ \label{kcsingle3}
\mathbf{p}_i &= \slashed{\mathbf{p}},
\end{align}
admit a solution, in the sense defined above (with real momenta and real, non-negative energies). In the above, $p_v$ and $\slashed{\mathbf{p}}$ are measured, whereas $p_i$ are four unknowns. Here, $m_T$ is defined by
\begin{gather}\label{mT}
m_T^2 \equiv (\alpha_v + \alpha_i)^2 = m_v^2 + m_i^2 + 2 (e_v e_i - \mathbf{p}_v \cdot \mathbf{p}_i).
\end{gather}

The proof is in two parts. First, we establish that any $(m_i,m_0)$ for which (\ref{kcsingle1}-\ref{kcsingle3}) have a solution is such that $m_0 \geq m_T (m_i)$. Second, we establish that (\ref{kcsingle1}-\ref{kcsingle3}) have a solution for $m_0 = m_T (m_i)$. 

For the first part, we have
\begin{align}
m_0^2 &= (p_v + p_i)^2 = p_v^2 + p_i^2 + 2p_v \cdot p_i \\
&= m_v^2 + m_i^2 + 2(E_v E_i -  \mathbf{p}_v \cdot \mathbf{p}_i -q_v q_i).
\end{align}
But since $E_v E_i - q_v q_i \geq e_v e_i$ (with equality at $E_v q_i = E_i q_v$), we have that $(\alpha_v + \alpha_i)^2 \leq (p_v + p_i)^2$, or, in other words, $m_T (m_i) \leq m_0$. 

For the second part, we need to show that the equations
\begin{align}\label{kcsingle4}
p_i^2 &= m_i^2, \\ \label{kcsingle5}
(p_i + p_v)^2 &= m_T^2 (m_i^2), \\ \label{kcsingle6}
\mathbf{p}_i &= \slashed{\mathbf{p}},
\end{align}
admit a solution. Equations (\ref{kcsingle6}) fix $\mathbf{p}_i$, and equation (\ref{kcsingle4}) fixes $E_i$ in terms of $q_i$. Equation (\ref{kcsingle5}) is satisfied by requiring $(p_i + p_v)^2 = (\alpha_v + \alpha_i)^2$, which, as we just learnt, requires $\frac{q_i}{E_i} = \frac{q_v}{E_v}$. Now for a given event, $\frac{q_v}{E_v}$ takes a value in $[-1,1]$, and as the remaining unknown $q_i$ varies in $\mathbb{R}$, $\frac{q_i}{E_i}$ takes all values in $[-1,1]$. Thus, equations (\ref{kcsingle4}-\ref{kcsingle6}) admit a solution.

So for a given event, $m_0 = m_T (m_i)$ defines the boundary of the allowed region in $(m_i,m_0)$. Given multiple events, the allowed region shrinks to the intersection of the allowed regions for each event. In the limit of arbitrarily many events, we obtain the extremal boundary given by the kink curve. Its explicit form was derived in \cite{Gripaios:2007is,Barr:2007hy}.
Here we simply quote the result for the special case where the visible system consists of a single, massless particle. The locus of the extremal boundary is given by
\begin{align}\label{singlemax1}
m_0^2 - \tilde{m}_0^2 &= m_i^2  -\tilde{m}_i^2, \; \mathrm{for} \; m_i \leq \tilde{m}_i, \\ \label{singlemax2}
\frac{m_0^2}{\tilde{m}_0^2} &=  \frac{m_i^2}{ \tilde{m}_i^2},  \; \mathrm{for} \; m_i > \tilde{m}_i.
\end{align}
These are simply straight lines in the space of mass-squareds.
\section{Identical pair decays and $m_{T2}$}
The analogue for identical pair decays and $m_{T2}$ of the argument just given has already been given in \cite{Cheng:2008hk}, and in any case follows as a corollary from our analysis of non-identical pair decays to be given below. Here, we point out that it can be immediately used to derive the form of the extremal boundary or kink curve, from the form of the maximal curve for a single decay (given by (\ref{singlemax1}-\ref{singlemax2}) for the special case of massless visible particles).

Indeed, for a single event, the $m_{T2}$ locus is given by the boundary of the region in $(m_i,m_0)$ for which the equations 
\begin{align}\label{kc1}
p_i^2 &= m_i^2, \\
p_i^{\prime2} &= m_i^{\prime 2}, \\
(p_i + p_v)^2 &= m_0^2, \\
(p_i^\prime + p_v^\prime)^2 &= m_0^{\prime 2}, \\ \label{kc2}
\mathbf{p}_i + \mathbf{p}_i^\prime &= \slashed{\mathbf{p}},
\end{align}
have a solution, with $m_i^{\prime}=m_i$ and $m_0^{\prime}=m_0$. The extremal boundary is given by the intersection of the allowed regions for all possible event configurations. But in considering all events, we permit any value of $\slashed{\mathbf{p}}$, such that the last equation (\ref{kc2}) can always be satisfied and becomes trivial.\footnote{This does not hold if we only consider a subset of events, for example those in which initial state or upstream momentum is forbidden, such that $\slashed{\mathbf{p}} + \mathbf{p}_v + \mathbf{p}_v^\prime = 0$.} But then the remaining equations decouple into those for the individual decaying systems. For these systems, the extremal curve has already been computed in \cite{Gripaios:2007is}. Since we have identical decays, the extremal curves for the two individual systems are the same, and this same curve is the extremal curve for the pair decay. By this argument, which we call the `decoupling argument', we thus find a simple proof of the results previously obtained in \cite{Barr:2007hy,Cho:2007dh}.
\section{Non-identical pair decays.}
The most general case is the one in which neither the parents nor the invisible daughters have common mass. The kinematic constraints are then given by
(\ref{kc1}-\ref{kc2}), but now with  $m_i^{\prime}\neq m_i$ and  $m_0^{\prime} \neq m_0$.
It is simple enough to derive the form of the extremal boundary of the allowed region in such a case. Just as for identical pair decays, we may invoke the decoupling argument. The extremal locus is then given simply by the individual extremal loci for the individual decays. Thus, in the four-dimensional space $(m_i,m_i^\prime,m_0,m_0^\prime)$, the extremal locus (which is a two-dimensional surface) simply factorizes into the product of the two one-dimensional curves in $(m_i,m_0)$ and $(m_i^\prime,m_0^\prime)$.
\begin{figure}
\begin{center}
\includegraphics[width=0.95\linewidth]{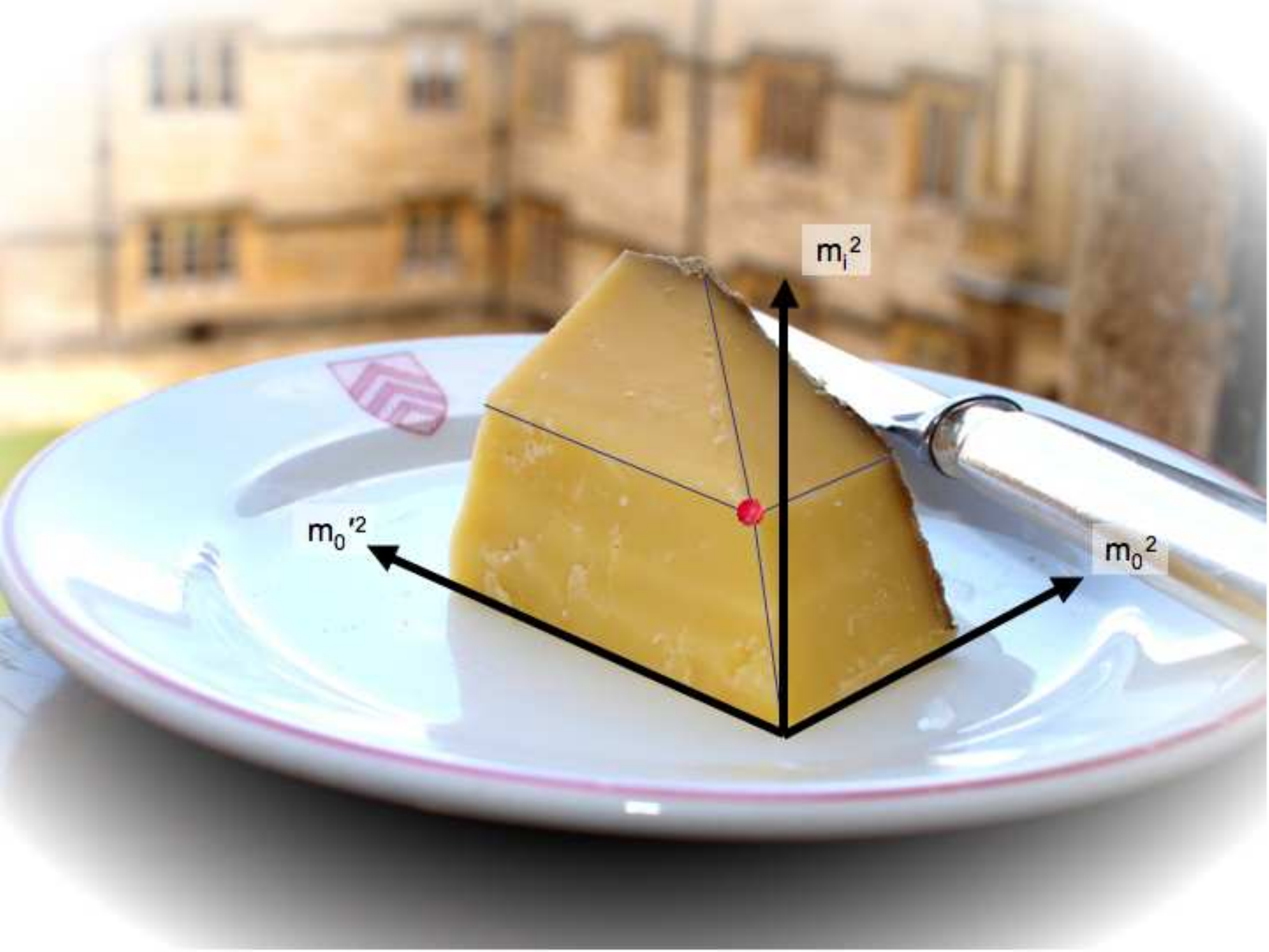}
\caption{\label{fig:intersections}
Representation of the bounding planes (visible faces) and the extremal allowed region (solid) for the case described in the text
with $\tilde{m}_i=\tilde{m}_i^\prime$, $m_i=m_i^\prime$, and $m_v=m_v^\prime=0$.
The vertex representing the true values of the masses is indicated with a red ball.
The origin of the axes is at the point $(m_0^2=\tilde{m}_0^2-\tilde{m}_i^2, m_0^{\prime2}=\tilde{m}_0^{\prime2}-\tilde{m}_i^2,m_i^2=0)$.
}
\end{center}
\end{figure}
Things become more interesting if we posit that either the parents or the daughters have a common mass. For example, if there exists a pair-produced, unique, stable dark-matter candidate, then the daughter particles in a pair decay will have common mass. Let us, for the sake of argument, consider this case in what follows. The allowed regions now occupy the 3-dimensional space parametrized by $(m_0,m_0^\prime,m_i = m_i^\prime)$. What is the form of the extremal boundary surface? By the decoupling argument, the extremal allowed region is given by the intersection of the extremal allowed regions for the individual decays. In the special case of massless visible particles, for example, the individual extremal allowed regions are bounded by
\begin{align}\label{unprime1}
m_0^2 - \tilde{m}_0^2 &= m_i^2  -\tilde{m}_i^2, \; m_0 \leq \tilde{m}_0, \\
\label{unprime2}
\frac{m_0^2}{\tilde{m}_0^2} &=  \frac{m_i^2}{ \tilde{m}_i^2},  \; m_0 > \tilde{m}_0,
\end{align}
for the unprimed system, and by
\begin{align}\label{prime1}
m_0^{\prime2} - \tilde{m}_0^{\prime2} &= m_i^2  -\tilde{m}_i^{2}, \; m_0^{\prime} \leq \tilde{m}_0^{\prime}, \\
\label{prime2}
\frac{m_0^{\prime2}}{\tilde{m}_0^{\prime2}} &=  \frac{m_i^{2}}{ \tilde{m}_i^{2}},  \; m_0^{\prime} > \tilde{m}_0^{\prime},
\end{align}
for the primed system. These both describe a surface in the space with co-ordinates $(m_0^2,m_0^{\prime2},m_i^2 = m_i^{\prime2})$. The extremal boundary for the pair decay is then given at each point $(m_0^2,m_0^{\prime2})$, by the surface that gives the smaller value of $m_i^2$. Let us assume that $\tilde{m}_0 > \tilde{m}_0^\prime$, without loss of generality. The extremal boundary surface is sketched in Figure \ref{fig:intersections}. 
The intersections of the four planes give rise to four creases in the extremal boundary surface, generalizing the kinks observed in extremal curves for identical pair decays. Furthermore, one can see that
various types of kink behaviour may arise by taking various two-dimensional slices through the three-dimensional space of masses. 

Consider, for example, the extremal curves of $m_0^\prime$ {\em vs.} $m_i$ obtained at fixed $m_0$. For $m_0^2>\tilde{m}_0^2$ there are two kinks --
the `usual one' at $(m_0=\tilde{m}_0^\prime,m_i=\tilde{m}_i)$ and a second on the upper part of the diagonal crease in Figure \ref{fig:intersections}; 
for $\tilde{m}_0^2-\tilde{m}_i^2 < m_0^2 < \tilde{m}_0^2$ there is a single kink on the lower part of the diagonal crease;
for $m_0^2 < \tilde{m}_0^2-\tilde{m}_i^2$ the allowed region is null.
The locus of the diagonal crease in $(m_0^2,m_0^{\prime2})$ is given by
\begin{align}\label{crease}
m_0^{\prime2}  -\tilde{m}_0^{\prime2} &= m_0^2  -\tilde{m}_0^2, \;  \tilde{m}_0^2 - \tilde{m}_i^2 \leq m_0^2 \leq \tilde{m}_0^2, \\
 \frac{m_0^{\prime2}}{ \tilde{m}_0^{\prime2}} &=  \frac{m_0^2}{ \tilde{m}_0^2},  \; m_0^2 > \tilde{m}_0^2.
\end{align}

A particularly striking kink is seen if one fixes $m_i^2$ and considers $m_0^{\prime2}$ as a function of $m_0^2$. There is then a right-angled kink coming from the diagonal crease. 

In the above, we have chosen to derive the extremal boundary directly, rather than relying on an explicit functional definition (like the original definition of $m_{T2}$). Nevertheless, an explicit functional definition is easily guessed. If we just have non-identical daughters, but common mass parents, we can simply use the original  $m_{T2}$ definition. But if the parents have distinct mass, the original $m_{T2}$ is no good, because its definition exploits the equality of the maximal values of $m_T$ and $m_T^\prime$. To maintain this equality, we propose a generalized definition of $m_{T2}$ for non-identical parent decays as
\begin{gather} \label{ratio}
m_{T2}^2 (m_i, m_i^\prime, m_0^\prime/m_0) \equiv \mathrm{min} \; \mathrm{max} ( \frac{m_0^\prime}{m_0} m_T^2,  \frac{m_0}{m_0^\prime} m_T^{\prime 2}) .
\end{gather}
Note that both $ \frac{m_0^\prime}{m_0} m_T^2 (\tilde{m}_i)$ and $\frac{m_0}{m_0^\prime} m_T^{\prime 2} (\tilde{m}_i^\prime)$ are bounded above by $m_0 m_0^\prime$.

It remains to show that this definition reproduces the boundary of the allowed region, event by event. As for single particle decays, there are two parts to the proof, which generalizes immediately from that given in \cite{Cheng:2008hk}.
First, we establish that any $(m_i,m_i^\prime,m_0, m_0^\prime)$ for which (\ref{kc1}-\ref{kc2}) have a solution is such that $m_0 m_0^\prime \geq m_{T2}^2 (m_i, m_i^\prime, m_0^\prime/m_0)$. This follows immediately from $m_T^2 (m_i) \equiv (\alpha_v + \alpha_i)^2 \leq (p_v + p_i)^2 = m_0^2$, from the corresponding inequality in the primed system, and from $m_{T2}^2 \leq \mathrm{max} ( \frac{m_0^\prime}{m_0} m_T^2, \frac{m_0}{m_0^\prime} m_T^{ \prime 2}) \leq  m_0 m_0^\prime$.

Second, we establish that (\ref{kc1}-\ref{kc2}) have a solution for $m^2_0 =  \frac{m_0}{m_0^\prime} m_{T2}^2 (m_i, m_i^\prime, m_0^\prime/m_0)$ and $m_0^{\prime 2} = \frac{m_0^\prime}{m_0} m_{T2}^2 (m_i, m_i^\prime, m_0^\prime/m_0)$. 
There are three possibilities to consider, arising from the three different ways in which values of $m_{T2}$ may arise \cite{Barr:2003rg}:
(i) the balanced case, with $m_{T2}^2 = \frac{m_0^\prime}{m_0} m_T^2 = \frac{m_0}{m_0^\prime} m_T^{\prime 2}$, 
(ii) the unbalanced case with $m_{T2}^2 = \frac{m_0}{m_0^\prime} m_T^{\prime 2} >   \frac{m_0^\prime}{m_0} m_T^2$, and 
(iii) the unbalanced case with $m_{T2}^2 = \frac{m_0^\prime}{m_0} m_T^2  >  \frac{m_0}{m_0^\prime} m_T^{\prime 2}$. 
In case (i), the solution of (\ref{kc1}-\ref{kc2}) is given by the $\mathbf{p}_i$ assigned by the minimization in the definition of $m_{T2}$ and the $q_i$ such that $\frac{q_i}{E_i} = \frac{q_v}{E_v}$, and similarly for the primed quantities. In case (ii), the solution has $\frac{q_i\prime}{E_i\prime} = \frac{q_v\prime}{E_v\prime}$; to find a suitable $q_i$, we note that, if we chose $q_i$ such that $\frac{q_i}{E_i} = \frac{q_v}{E_v}$, we would obtain $(p_i + p_v)^2 = m_T^2 <  \frac{m_0}{m_0^\prime} m_{T2}^2$, whereas if we chose $q_i \rightarrow \infty$, we would find $(p_i + p_v)^2 \rightarrow \infty $. Since $(p_i + p_v)^2$ is a continuous function of $q_i$ on $\mathbb{R}$, there must, by the intermediate value theorem, exist values of $q_i$ such that $(p_i + p_v)^2 =  \frac{m_0}{m_0^\prime} m_{T2}^2$, as required. A similar argument applies to case (iii).

Thus we have proven that the explicit definition of $m_{T2}$ for non-identical decays in (\ref{ratio}) reproduces the kinematic boundary surface. Interestingly, it would appear that, in the case of distinct daughter and parent masses, the boundary for each event is given by a three-dimensional hypersurface in the four-dimensional space of $(m_0, m_0^\prime, m_i , m_i^\prime)$, whose locus is $m_0 m_0^\prime = m_{T2}^2 (m_i, m_i^\prime, \frac{m_0^\prime}{m_0})$.
By contrast, the {\em extremal} boundary in this case, whose form we derived at the beginning of this Section, is a two-dimensional surface in four dimensions.
\section{Combinatorics and $m_{T\mathrm{Gen}}$}
In the real world of experiment, one must also face the fact that in pair decays there will be combinatoric ambiguities.
For example, in identical pair decays, there are (at least) two copies of each visible particle in the final state, and one does not know which decay to assign them to. With $2n$ visible particles, this results in a $2^{(n-1)}$-fold ambiguity. As a result, the visible momenta in equations (\ref{kc1}-\ref{kc2}) are themselves ambiguous. In the presence of such an ambiguity, the allowed kinematic region is given by the union of the allowed regions obtained by considering all possible branch assignments. Equivalently, the boundary of the allowed region is obtained by taking the minimal $m_{T2}$ curve with respect to the different combinatoric assignments. This corresponds to the $m_{T\mathrm{Gen}}$ variable defined in \cite{Lester:2007fq}. 

Similar combinatoric ambiguities can arise in the presence of upstream
or initial state radiation. Again, to find the allowed kinematic region, one simply takes the union of regions obtained by considering all possible assignments, which is equivalent to the $m_{T2}$-based prescription given in \cite{Alwall:2009zu}.
\section{Derived observables and one-dimensional distributions}\label{distros}
The kinematic boundary hypersurface for a decay topology gives a complete picture of the kinematic constraints coming from an event. In principle, it can be generated from experimental data, but in practice, it is not clear how this will be achieved. More likely is that experimentalists would prefer to generate one-dimensional distributions of specific observables, which they can then compare with the results of numerical simulations. 

In order to do so, one would like to understand how to `translate' the kinematic boundary plot into observables. This can certainly be achieved if one has determined all but one of the unknown masses. One simply needs to isolate the pertinent observable. For example, in an identical pair decay, if one knows the mass of the invisible daughter, the 
$m_{T2} (\tilde{m}_i)$ distribution can be used to measure the mass of the parent. Conversely, one could imagine that one knew the mass of the parent, and wished to extract the mass of the invisible daughter. Clearly what one needs in this case is an observable derived from the inverse function of $m_{T2} (m_i)$. 

To derive an explicit expression for the inverse function of $m_{T2}$, consider first a single decay and $m_T$.
We claim that the inverse of $m_{T}$ for an event is given by\footnote{We note that the inverse of $m_T$ may also be useful on its own for hadron collider mass measurements. For example, one could measure the charged Higgs mass in decays $t \rightarrow H^+ b$, followed by the decay $H^+ \rightarrow \tau \nu$ with an invisible daughter. In fact, the variable defined in \cite{Gross:2009wi} is precisely the inverse transverse mass of the $t bH^+$ system, with parent $t$ and invisible daughter $H^+$.}
\begin{align}
(m_{T}^2)^{-1} (m_0) \equiv (\alpha_0 - \alpha_v)^2, \; m_0 \geq m_v.
\end{align}
(The condition $m_0 \geq m_v$ is added to ensure that $m_T (m_i)$ is surjective, such that $m_{T}^{-1}$ exists.) The reader may easily check explicitly that
$m_{T}^{-1} (m_T (m_i)) = m_i$. 

For pair decays, the appropriate definition is
\begin{align}
m_{T2}^{-1} (m_0) \equiv \mathrm{max \; min} ( m_{T}^{-1}, m_{T}^{\prime -1}).
\end{align}
Note that, rather than minimizing over momentum assignments, we now maximize over them, and, rather than taking the larger of the two observables, we now take the smaller of the two. 
To demonstrate explicitly that this is the inverse, it suffices to show that both $m_{T2}$ and its inverse give rise to the same kinematic boundary for an event.

We note that the inverse of $m_{T2}$ can be applied directly to measure the mass of a common invisible daughter, even if the parent masses are not identical, provided the parent masses are themselves known.
\begin{figure}
\begin{center}
\includegraphics[width=0.8\linewidth]{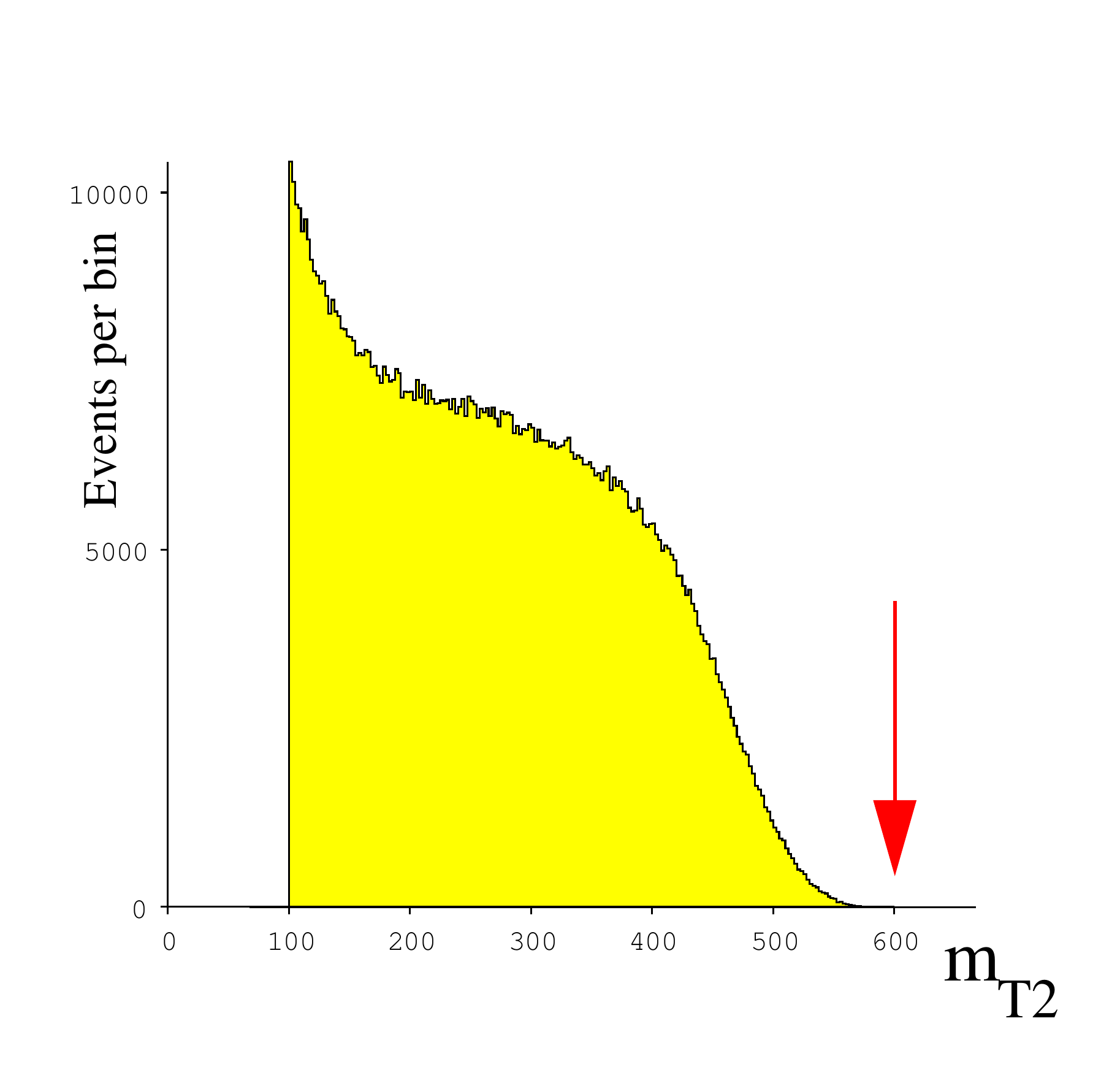}
\caption{\label{fig:mt2dissimilarmasses}
Distribution of the usual $m_{T2}$ for $m_0=400$, $m_0^\prime=600$,
$m_i=100$, {\em i.e.}~for a case in which the parent particles have different
masses.  The distribution has a kinematic endpoint at the mass of the heavier
particle (indicated by the red arrow) but it will be noted that this
endpoint is not very useful, as the density of states near the end point
is low.
}
\end{center}
\end{figure}

Finally, let us ask what are the appropriate observables for non-identical decays? If one knew, say, the two daughter masses as well as the {\em ratio} of the two parent masses, one could simply invoke the observable defined in (\ref{ratio}) to determine the {\em product} of the parent masses. If, alternatively, one knew the two daughter masses as well as the lighter parent mass, one could use the usual $m_{T2}$ definition. Indeed, it is easy enough to see that $m_{T2}$ is bounded above by the mass of the heavier parent. Unfortunately, the $m_{T2}$ distribution has a very poor endpoint behaviour, as we illustrate in Figure~\ref{fig:mt2dissimilarmasses}. Experimentally, such fine edges are likely to be difficult to catch.

Since the distribution of the ratio variable (\ref{ratio}) has much better end-point behaviour 
-- it has thick edges, especially when the correct value of the ratio is used --
the pragmatic choice may be to hypothesise different values for the ratio, and to compare
the experimental data to template distributions of (\ref{ratio}) for each value of $\frac{m_0^\prime}{m_0}$.

\section{Mass measurement via the matrix element}
If one knew (or guessed) explicitly the Lagrangian, one could hope to measure the
masses directly using the likelihood and the matrix element. In
practice this reduces to kinematic constraints plus constraints from
parton distribution functions (PDFs), plus a small dependence on
dynamics (spins/couplings etc), which in any case one would presumably
not want to trust to begin with. The arguments of Cheng and Han tell
us that all of the information from kinematics is encoded in $m_{T2}$,
which cuts off the lower right region of the $(m_i,m_0)$ plane. The PDFs limit the
allowed mass of the parent, cutting off the upper part of the
plane. So one can already see, roughly speaking, what the negative likelihood
contours of a matrix element method will look like.  They will
describe a narrow gully lying along the boundary of the extremal
$m_{T2}$ curves.  
Evidence is beginning to emerge which supports this conjecture \cite{JohanIpmu}.
\section{Summary}

Cheng and Han's interpretation of the function $m_{T2}$ as the kinematic boundary between 
allowed and disallowed regions of mass space is a powerful one.
It has allowed us to prove more elegantly the known results for the extremal boundaries
for single-particle and identical pair decays.
Its generalization has allowed us to prove new results for non-identical pair decays 
and for complex pair-decay topologies with indistinguishable particles in the final state.

We have constructed three explicit examples of bounding functions
that perform roles similar to $m_T$ and $m_{T2}$, but with differing assumptions. 
The first is a generalization of $m_{T2}$ that is appropriate when parents with different masses decay to
equal-mass invisible daughters -- a case which will be of particular interest at the LHC.
The other generalizations are the inverse functions $m_T^{-1}$ and $m_{T2}^{-1}$
which require the parent particle mass as a parameter, and
which then provide the extremal bound on the invisible daughter (WIMP) mass.

As well as providing a mass-determination method in their own right,
such variables encode the kinematic part of the likelihood function. 
This means that insights gained from their construction can inform one's interpretation
of mass determinations using the full matrix element -- where such calculations are computationally tractable.
This final test will show whether it is safe to neglect the effects of spin,
determine the character of the creases, and get the desired results by using the boundary.

\acknowledgements
We are grateful to the IPMU, Tokyo for hosting the workshop where the discussions leading to this paper were initiated.
BMG and CGL also thank the Les Houches workshop on `Physics at TeV
Colliders 2009' for hospitality. We thank colleagues in the Cambridge Supersymmetry Working Group (Chris Cowden in particular) and the Dalitz Institute, Oxford for useful discussions.
AJB is supported by an Advanced Fellowship from the UK Science and Technology Facilities Council, and by Gruy\`ere from 
the benefice of Walter de Merton.

\bibliography{paper}
\end{document}